# Optical biosensor based on a photonic crystal with a defective layer designed to determine the concentration of SARS-CoV-2 in water

I.M. Efimov[1], N.A. Vanyushkin[1], A.H. Gevorgyan[1], S.S. Golik[1,2]

[1]School of Natural Sciences, Far Eastern Federal University, 10 Ajax Bay, Russky Island, 690922 Vladivostok, Russia

[2]Institute of Automation and Control Processes, Far East Branch, Russian Academy of Sciences, Vladivostok, 690041 Russia

**Abstract**

We propose a new optical biosensor based on a photonic crystal with a defect layer, which can determine the SARS-CoV-2 concentration in water by the defect mode shift. Two models of the dependence of the refractive index of the defect layer on the concentration of the pathogen in water were considered. The optimal parameters for the photonic crystal in our device and the optimal thickness of the defect layer were determined. It was also demonstrated that in the presence of absorption in the investigated structure it is much more advantageous to work in the reflection mode compared to the transmission mode. Finally, the wavelength dependence of the defect mode on the SARS-CoV-2 concentration was obtained and the sensitivity of the sensor was determined.

## 1.    Introduction

Pathogens are various microorganisms, such as viruses and bacteria, which can cause damage to the host organism. Throughout history, pathogens have accompanied humanity and the emergence of new species of infectious pathogens has caused various epidemics [1-3]. Many scientists of the last century saw antibiotics as a solution to the problem, but antibiotic-resistant pathogens, which significantly complicate treatment and recovery of patients, are also a real threat. The next step in solving this problem is the timely detection and isolation of pathogen vectors.

The rapid development of genetics and biochemistry has led to the expansion of various medical and biomedical tools to detect various pathogens with great accuracy. There are now many methods for detecting pathogens, including SARS-CoV-2. The best known and most used identification procedure is based on real-time reverse transcription polymerase chain reaction (RT-PCR) [4]. However, this diagnosis requires advanced laboratory testing, expensive equipment, and experienced personnel. Therefore, it is necessary to simultaneously simplify and automate the process of pathogen detection. Nowadays, various optical biosensors are being actively developed, because they are highly sensitive to changes in pathogen concentration, can operate in different modes, such as transmittance, reflection and absorption of light [1-14].

Currently, there are many types of optical biosensors. They include biosensors based on plasmon resonance spectroscopy [4]. Plasmons are collective fluctuations in the charge density of a free electron gas. The simplest system in which surface plasmons can be excited is a metal-dielectric boundary [4, 12-15]. There are also biosensors based on the Raman effect, inelastic scattering of optical radiation on molecules of matter, accompanied by a significant change in the frequency of radiation. The number and location of lines appearing is determined by molecular structure of the substance [15-18]. Another type is fluorescence-based biosensors [7]. Their mechanism can be based on fluorescence quenching, fluorescence amplification or resonance fluorescence energy transfer.

In recent years, biosensors based on photonic crystals (PCs), including those with a defect in the structure, have been of great interest [18-24]. PCs are a periodic structure of layers with different refractive index. PCs have the unique property of having a certain range of frequencies, called the photonic bandgap (PBG), in which an electromagnetic wave cannot propagate through the PC [25]. In practice, this means that if radiation with a wavelength inside the PBG is incident on the PC, it experiences strong reflections from the PC. Thus, the PC can act as a mirror or an optical filter. If a defective layer is added to the periodic structure of the PC, the periodicity of the structure is violated, which leads to changes in the transmission and reflection spectra in the entire region. This manifests itself in the appearance of a narrow band of transmittance inside the PBG, which is called a defect mode (DM). The position and shape of the DM depends on the parameters of the defect layer, such as the thickness and refractive index of the defect layer. It is this property that underlies our biosensor. In this work, the defect layer is a layer composed of water with the addition of SARS-CoV-2. Changing the concentration of the pathogen leads to a change in the refractive index of the defect layer, which in turn changes the position of the defect mode, which can be detected by the sensor. The SARS-CoV-2 concentration in the defect layer is determined by this deviation.

## 2. The Theory

The transverse structure of the PC we are considering is shown in Figure 1.

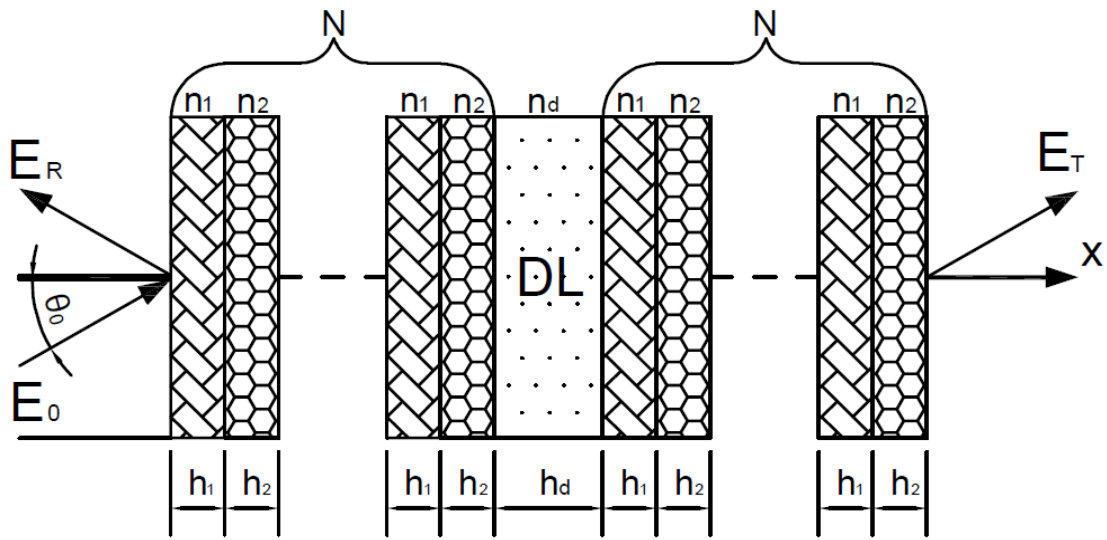

Figure 1. Schematic diagram of the analyzed structure

$E_R$ – Incident wave;

$E_0$ – Reflected wave;

$E_T$ – Transmitted wave;

$N$ – Number of unit cells;

$\theta_0$ – Angle of incidence;

$DL$ – Defect Layer.

Our structure consists of a defective layer sandwiched between two perfect PCs, each of which consists of $N$ periodic cells. Each cell is a pair of layers with thickness $h_{1,2}$ and refractive index $n_{1,2}$. The refractive indices were determined by the materials of the layers. To select suitable materials that can be used as cell layers, the following conditions were set: the materials should provide a high difference between the refractive indices of the $n_1$ and $n_2$ layers, be relatively inexpensive and easy to manufacture, and transparent in the spectral range of interest. The thicknesses of the layers were determined from the quarter-wave criterion $Re[n_1(\lambda_b)] h_1 = Re[n_2(\lambda_b)] h_2 = \lambda_b/4$, where $\lambda_b$ is the center wavelength PBG. When this condition is met, we get the maximum width of PBG.

Next, let us consider the defect layer. The defect layer is a medium of host material with particle inclusions. In our case the host material is water, and the inclusions are SARS-CoV-2 pathogens. Two models were used to determine the refractive index of the defect layer through the parameters of the constituent substances: the Maxwell-Garnett effective medium approximation (EMA) and the simple proportional dependence of the dielectric permittivity. EMA refers to analytical or theoretical medium modeling that describes the macroscopic properties of composite materials [26-27]. This method is applicable if the macroscopic system is homogeneous, and the sizes of all particles are much smaller than the wavelength [26-27]. The calculation of the dielectric permittivity of the composite by EMA is represented by the equation:

$$\varepsilon_m = \varepsilon_M \frac{2\delta_i(\varepsilon_i - \varepsilon_M) + \varepsilon_i + 2\varepsilon_M}{2\varepsilon_M + \varepsilon_i - \delta_i(\varepsilon_i - \varepsilon_M)}, \quad (1)$$

where $\varepsilon_m$ is effective permittivity of the composite, $\varepsilon_i$ is dielectric constant of inclusions, $\varepsilon_M$ is dielectric constant of the host material, $\delta_i$ are volume fraction of the inclusions.

The proportional dependence is based on the assumption that the dielectric permittivity depends linearly on the volume fractions of the host material and inclusions. The dependence is represented by the equation:

$$\varepsilon_m = \varepsilon_M (1 - \delta_i) + \varepsilon_i \, \delta_i, \quad (2)$$

In this paper we used the transfer-matrix method [28-30] to calculate the transmission and reflection spectra of the structure under study. The transfer matrix for the j-th layer in the structure can be written as:

$$M_j = \begin{pmatrix} \cos k_j h_j & \frac{-i}{p_j} \sin k_j h_j \\ -i p_j \sin k_j h_j & \cos k_j h_j \end{pmatrix}, \quad (3)$$

where $k_j = \frac{2\pi}{\lambda} n_j \cos \theta_j$, $\theta_j$ is the angle of incidence in the j-th layer, which is defined from Snell's law as: $\theta_j = \cos^{-1}\sqrt{1 - \frac{n_0^2 \sin^2 \theta_0}{n_j^2}}$, $p_j = n_j \cos \theta_j$, $n_0$ is the refractive index of the external medium to the left of the PC, $\theta_0$ is angle of incidence.

After that, the matrix $M$ of one unit cell in the periodic part of the structure is obtained by multiplying the transfer matrices $M_j$ of the layers in the cell:

$$M = M_2 M_1. \quad (4)$$

Since M is unimodular ($det\, M = 1$), the power matrix $M^N$, for N period structure, can be written from Chebyshev polynomials [28-29] as:

$$M^N = \begin{pmatrix} M_{11} U_{N-1}(X) - U_{N-2}(X) & M_{12} U_{N-1}(X) \\ M_{21} U_{N-1}(X) & M_{22} U_{N-1}(X) - U_{N-2}(X) \end{pmatrix}. \quad (5)$$

$U_N$ denotes Chebyshev polynomials of the second kind and they are given by:

$$U_N(X) = \frac{\sin[(N+1)\cos^{-1} X]}{\sqrt{1-X^2}}, X = \frac{1}{2}(M_{11} + M_{22}). \quad (6)$$

Finally, the transfer matrix of the whole structure has the following form:

$$m = M^N M_d M^N = \begin{pmatrix} m_{11} & m_{12} \\ m_{21} & m_{22} \end{pmatrix}. \quad (7)$$

where $M_d$ is transfer matrix of the defective layer.

Carrying out some mathematics then the transmission and reflection coefficients are given as:

$$t = \frac{2p_0}{(m_{11}+p_f m_{12})p_0+(m_{21}+p_f m_{22})}, \quad (8)$$

$$r = \frac{(m_{11}+p_f m_{12})p_0-(m_{21}+p_f m_{22})}{(m_{11}+p_f m_{12})p_0+(m_{21}+p_f m_{22})}, \quad (9)$$

Energy transmission and reflection coefficients:

$$T = \frac{p_f}{p_0}|t^2|, \quad (10)$$

$$R = |r^2|, \quad (11)$$

where $0, f$ is indices denote the parameters of the environment bordering the PC on the left and on the right, respectively.

## 3. The Results and Discussion

In our work, silicon oxide ($SiO_2$) (first layer) and silicon (Si) (second Layer) layers were used as materials for the PC unit cell. The dielectric permittivity of Si [31] and $SiO_2$ [31] as a function of wavelength is shown in Figure 2.

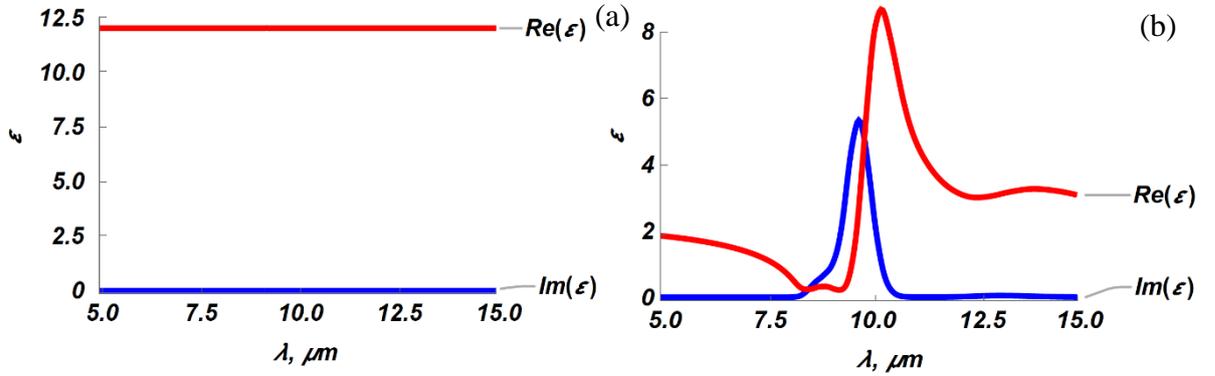

Figure 2. Spectrum of dielectric permittivity of (a) Si and (b) $SiO_2$

In the following, we considered the case of normal incidence ($\theta_0=0$) and each perfect PC in our structure had $N = 8$ unit cells. The external environment is vacuum ($n_0=0$).

Now consider the defect layer. The dielectric permittivity of the matrix [32] and inclusions [33-34] is shown in Figure 3.

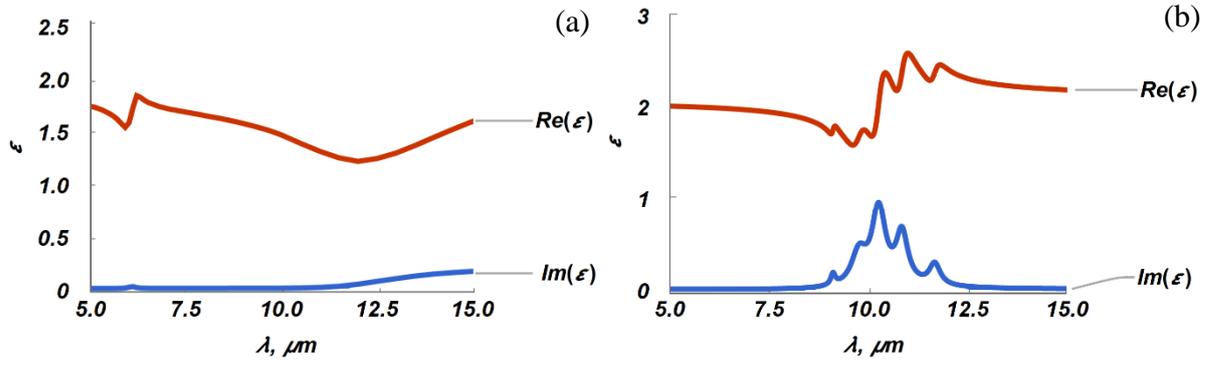

Figure 3. Spectrum of dielectric permittivity of $H_2O$ (a) and SARS-CoV-2 (b)

Figure 4a shows the difference in the permittivity of the defective layer $Re(\Delta\varepsilon) = Re(\varepsilon_{Covid}) - Re(\varepsilon_{H_2O})$, and in Figure 4b the sum of the imaginary part of the refractive indices of all materials of our structure $Im(\sum \varepsilon) = Im(\varepsilon_{Covid}) + Im(\varepsilon_{H_2O}) + Im(\varepsilon_{SiO_2}) + Im(\varepsilon_{Si})$.

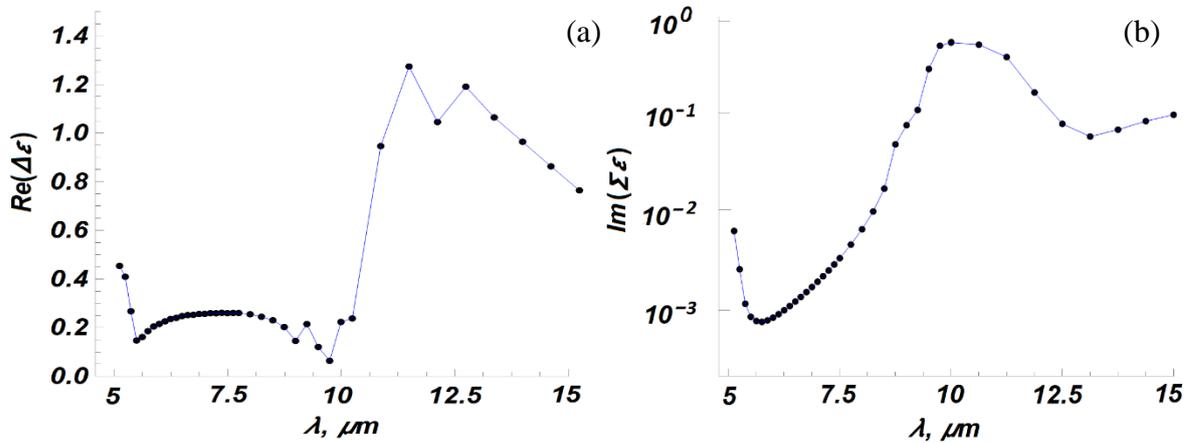

Figure 4. Comparative characteristics of the components of the PC

Obviously, the higher the difference in the refractive indexes of the defect layer components, the higher the sensitivity to changes in the pathogen concentration, but at the same time, as it will be shown below, high imaginary component values of each materials reduce the DM amplitude. Therefore, we were faced with the choice of which spectral range to work in. In the 10-15 μm range, the defect layer components have a high refractive index contrast, but all materials except Si strongly absorb radiation. On the other hand, In the 5-8 μm range, the situation is the opposite.

As noted above, models (1)-(2) were used to determine the refractive index of the defective layer. Figure 5 shows the dependences of the real and imaginary components of the refractive index as a function of the pathogen concentration at wavelength $\lambda = 5$ μm, calculated using EMA.

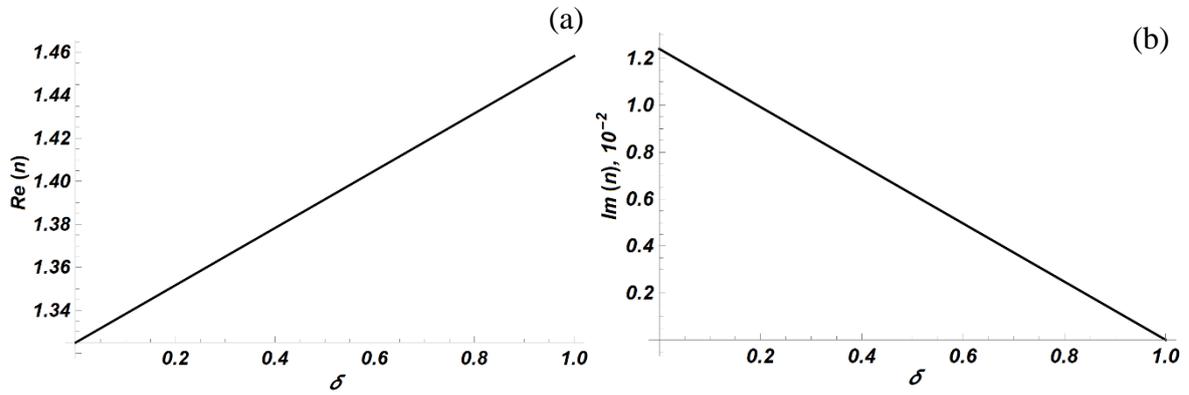

Figure 5. Dependences of the refractive index of the defect on SARS-CoV-2 volume fraction, at a wavelength of $\lambda = 5$ μm, (a) Real part, (b) Imaginary part

Figure 6 shows the difference between the refractive indices calculated by the EMA method and the proportional relationship at different wavelengths, $\Delta n = Abs(Re(n_{EMA} - n_P)/Re(n_{EMA}))$.

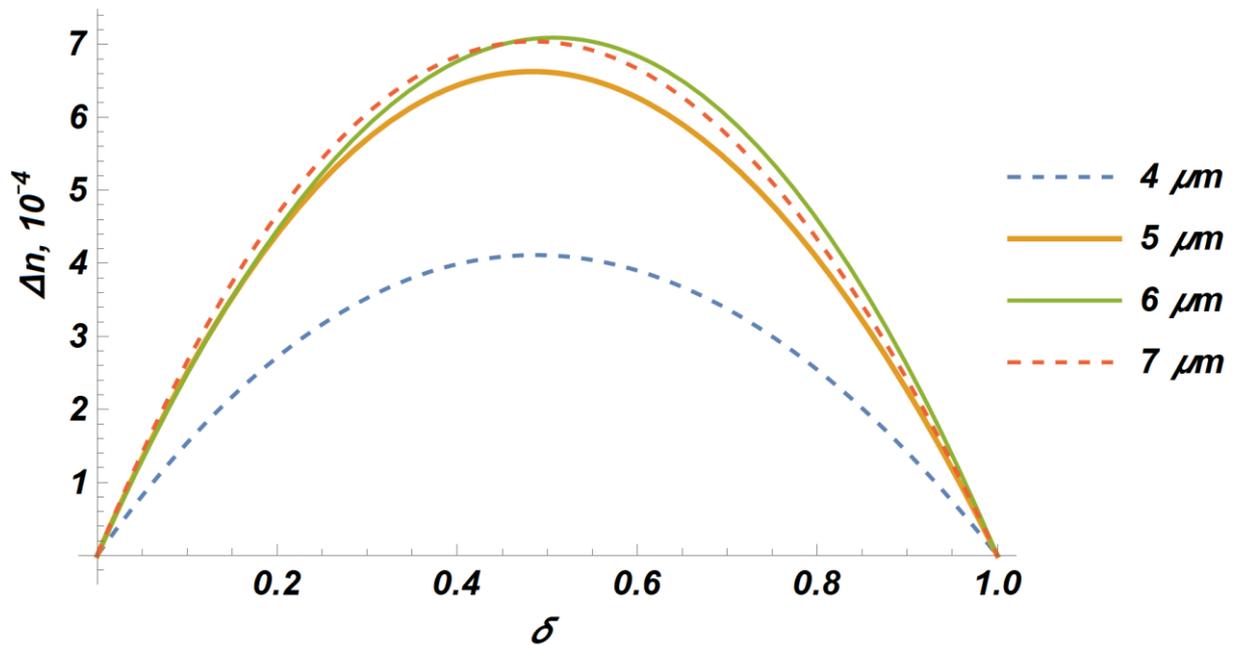

Figure 6. Difference in calculated values of the defect's real part of refractive index between EMA and the proportional dependence

As can be seen from Fig. 6, the maximum difference between the two methods is achieved at equal pathogen and water concentrations. [27]. It is also worth noting that, in principle, the SARS-CoV-2 concentration would be close to 0 in the real situation, and by examining Figure 6 we can see that the difference in the refractive index of the defect layer at extremely small concentrations is insignificant. Nevertheless, further in our work we will use the EMA method.

At the beginning of our study, we considered the wavelength range from 5 μm to 15 μm in which the optical properties of SARS-CoV-2 are known [33-34]. Figure 7 shows a series of R and T spectra at different PBG positions without pathogen particles in the defect layer. Spectra series a, b shows the negative influence of high $SiO_2$ absorption, spectra series c, d shows the negative influence of $H_2O$ absorption, and series e, f shows the variant with minimal negative influences. DMs are marked by black arrows. Blue dashed lines mark the PBG boundaries

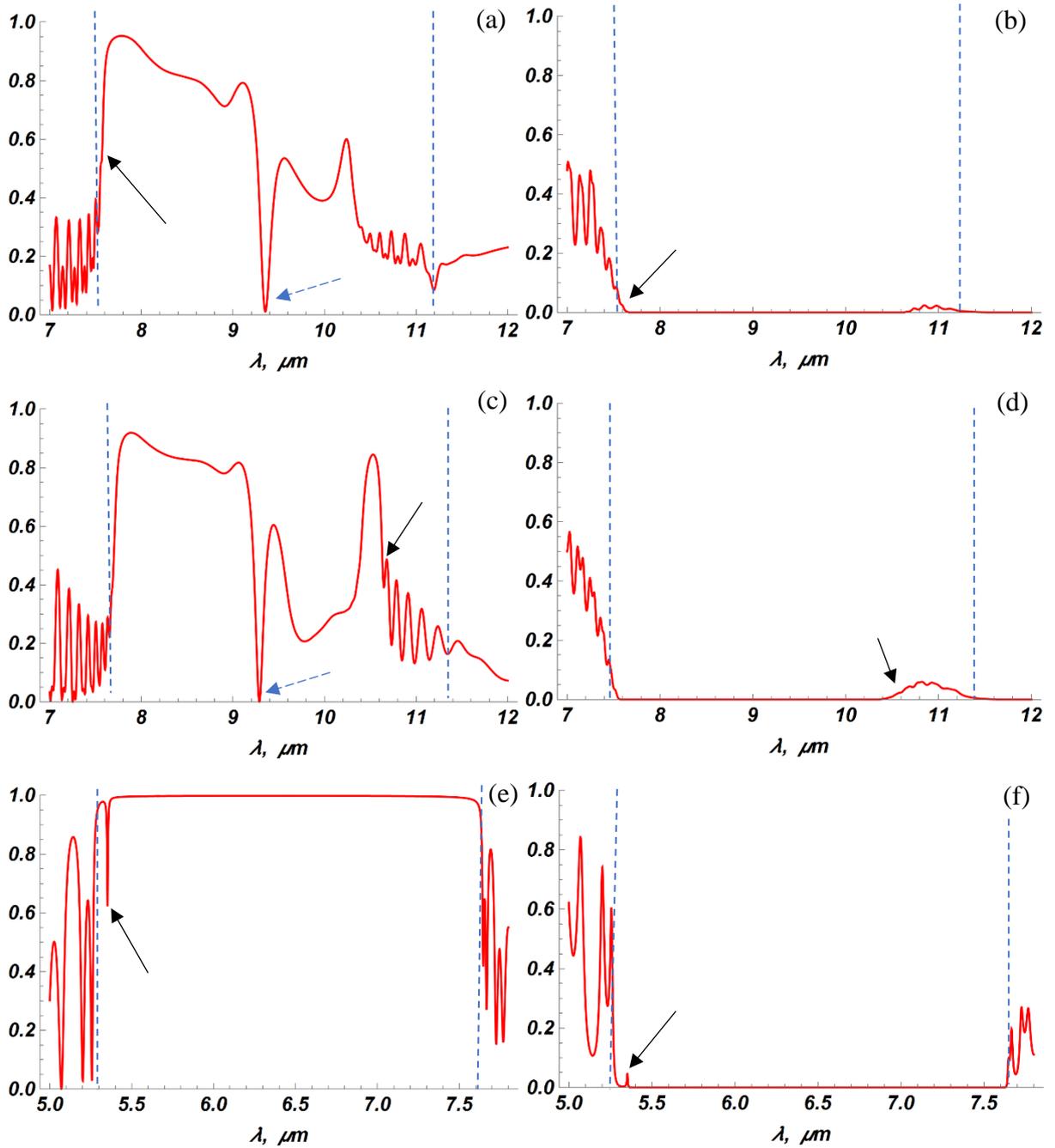

Figure 7. Reflectance and transmittance spectra at $\delta_i = 0$ with different parameters: (a, b) $h_1 = 3.00$ μm, $h_2 = 1.10$ μm, $h_d = 0.30$ μm; (c, d) $h_1 = 3.50$ μm, $h_2 = 1.41$ μm, $h_d = 1.50$ μm; (e, f) $h_1 = 1.30$ μm, $h_2 = 0.490$ μm, $h_d = 0.15$ μm.

Further in our study, we consider the range from 5 μm to 7.5 μm. This range is chosen because minima of $Im(n)$ of all components of the structure, which is responsible for absorption and reduces the amplitude of the DM are observed in this range. This effect can be seen in Figures 7. The main contribution to absorption at the selected wavelengths is made by water, so it is optimal to choose the operating wavelength near a water absorption minimum. We chose the water absorption minimum near 5.3 μm. When comparing a series of spectra with and without a defective layer, it can be observed that when a defective layer is added, not only DM appears, but also the splitting of the edge modes outside the PBG occurs (see also [35]). In Figures 7a, c a deep reflection minimum can be observed at 9.3 um (blue dash arrow). This minimum is due to high absorption by $SiO_2$ (see Figure 2b). Figure 8 shows the R spectra in the case in Figure 7e with and without the defective layer in the vicinity of the left PBG boundary.

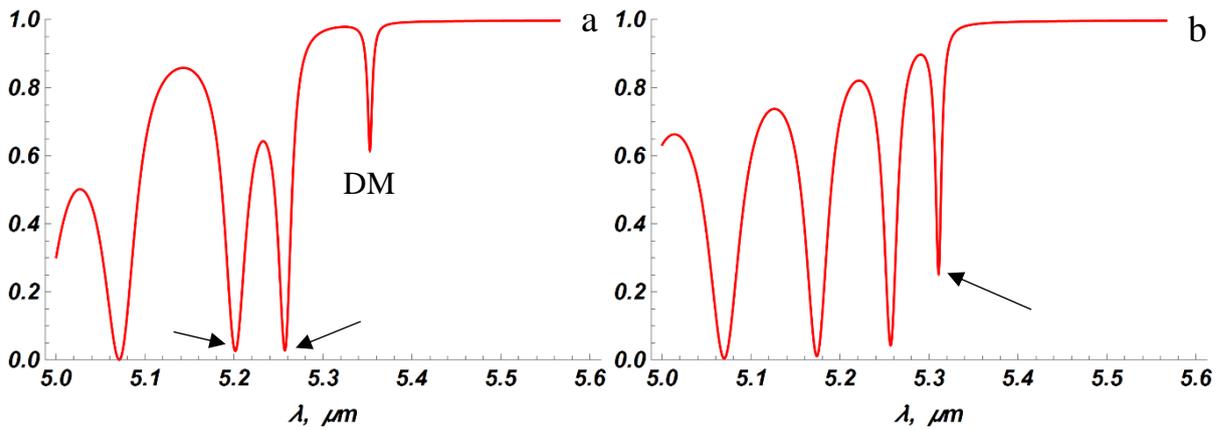

Figure 8. Reflection spectra near the left PBG boundary with different thickness of the defect layer: (a) $h_d = 0.15$ µm; (b) $h_d = 0$ µm. The other parameters: $h_1 = 1.30$ µm, $h_2 = 0.49$ µm, $\delta_i = 0$.

It is also worth noting the appearance of a slight shift of the PBG boundary. In the case of adding a defective layer $h_d = 0.15$ µm, the left boundary of the PBG has experienced a blue shift by the value of $\Delta\lambda = 0.06$ µm.

To view the full picture of the influence of the defect layer thickness on the DM, the reflection and transmittance spectra of our structure depending on the thickness of the defect layer were plotted in Figure 9. For clarity, the reflection spectrum is depicted as $1 - R$, which is equivalent to the sum of transmission and absorption $T + A$.

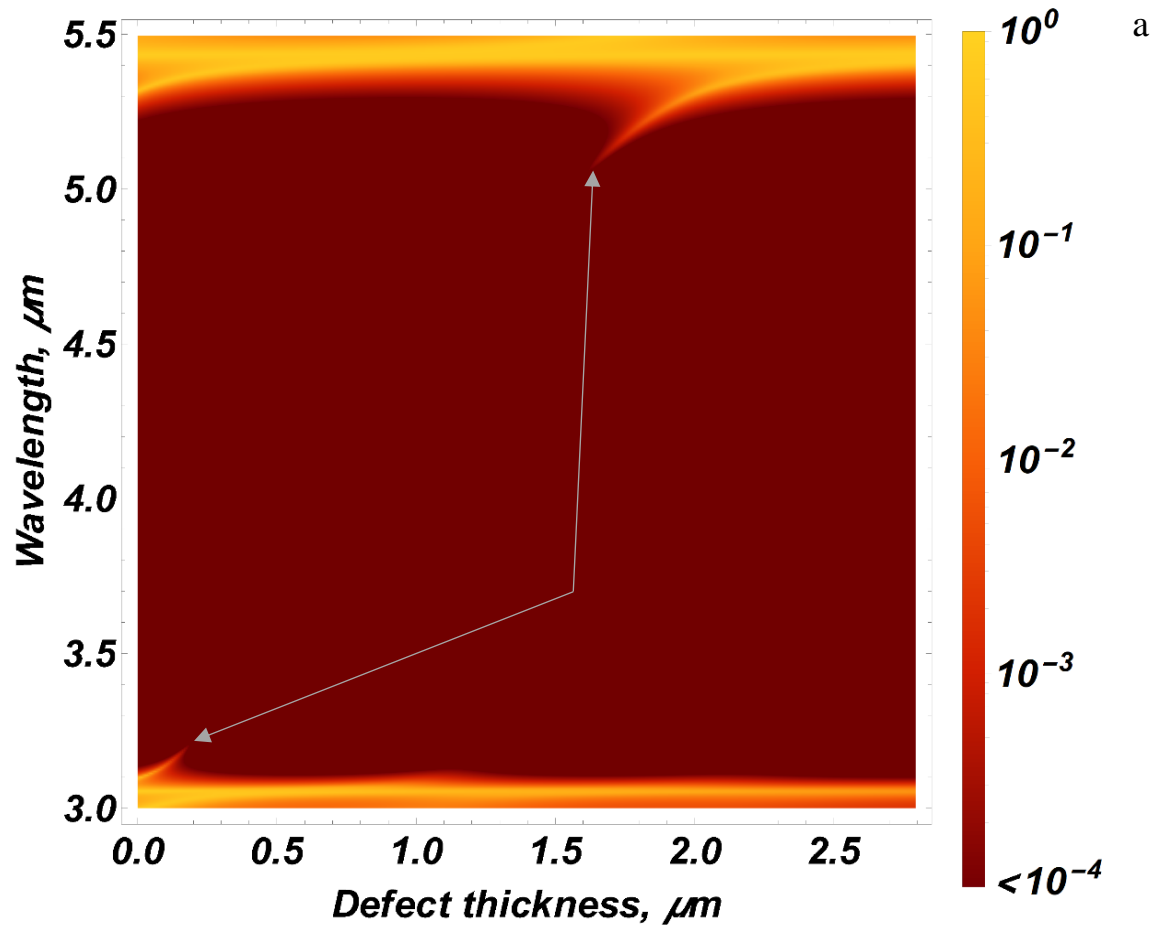

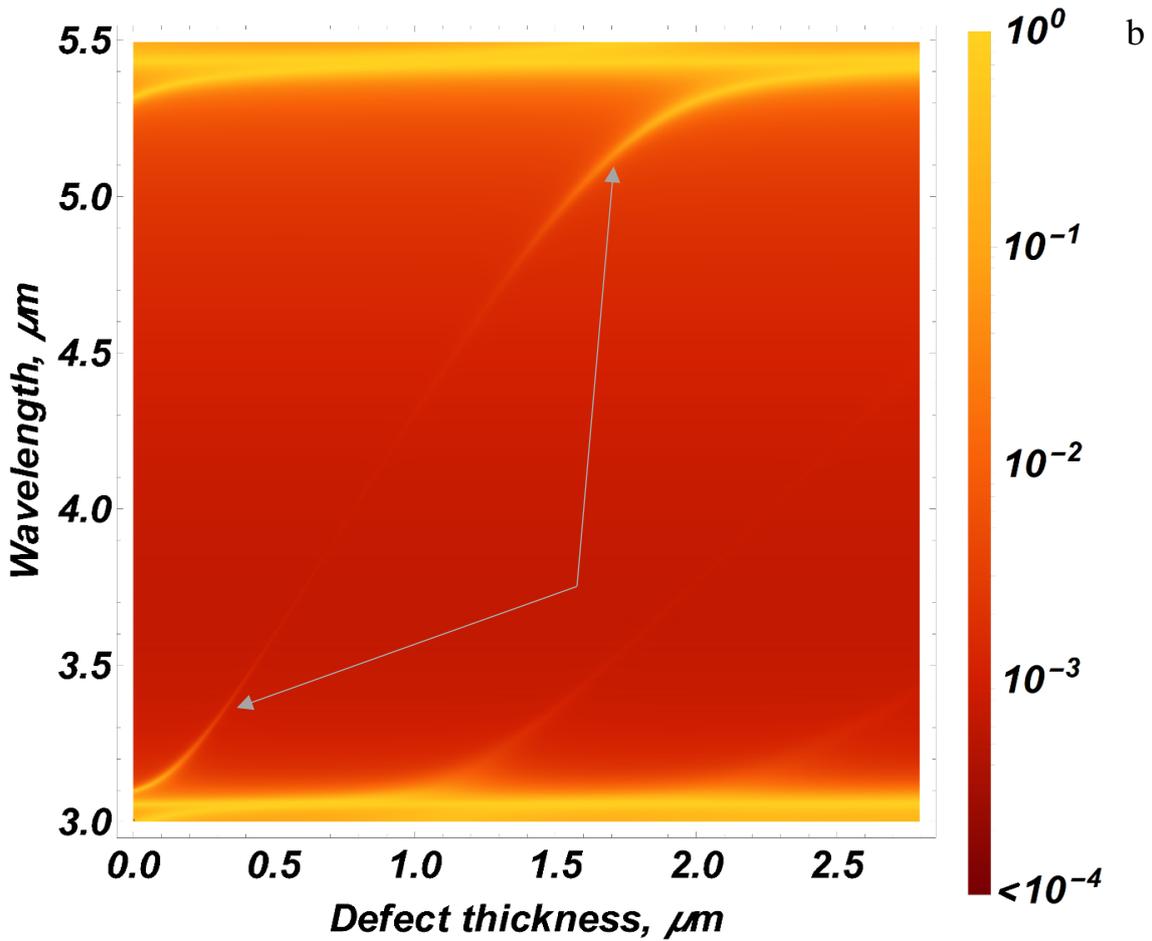

Figure 9. Transmittance (a) and reflectance (b) spectra with varying thickness of the defect layer, $h_1 = 0.700$ μm, $h_2 = 0.268$ μm

The arrows indicate the first DM, which has the highest sensitivity. As can be seen, the defect mode appears stronger at the boundaries of the PBG, especially at the longwave one, while in the center of the PBG the influence of the defect is practically not observed. This can be explained by the strong influence of absorption in the structure on the amplitude of the defect mode near the center of the PBG. One can also note the higher amplitude of DM on the reflection spectrum. The sensitivity of the defect mode to changes in the optical thickness of the defect $h_d n_d$ is maximum at the center of the PBG (see the slope of the curve showing the mode position in Figure 9), but as we approach the center of the PBG the amplitude of the defect mode, as noted above, drops, so we sought a compromise between sensitivity and amplitude. Based on Figures 9a and 9b, only the reflection spectra will be considered, due to higher amplitude of DM.

Based on the reasoning above, let us consider the case near the water absorption minimum at wavelength $\lambda = 5.30$ μm. As it was shown earlier, approaching the center of the PBG increases the sensitivity, so we tried to achieve the highest sensitivity while maintaining the minimum DM amplitude at the level of at least 0.1 [36]. Figure 10 shows a series of reflection spectra with different defect layer parameters. DMs are marked by black arrows.

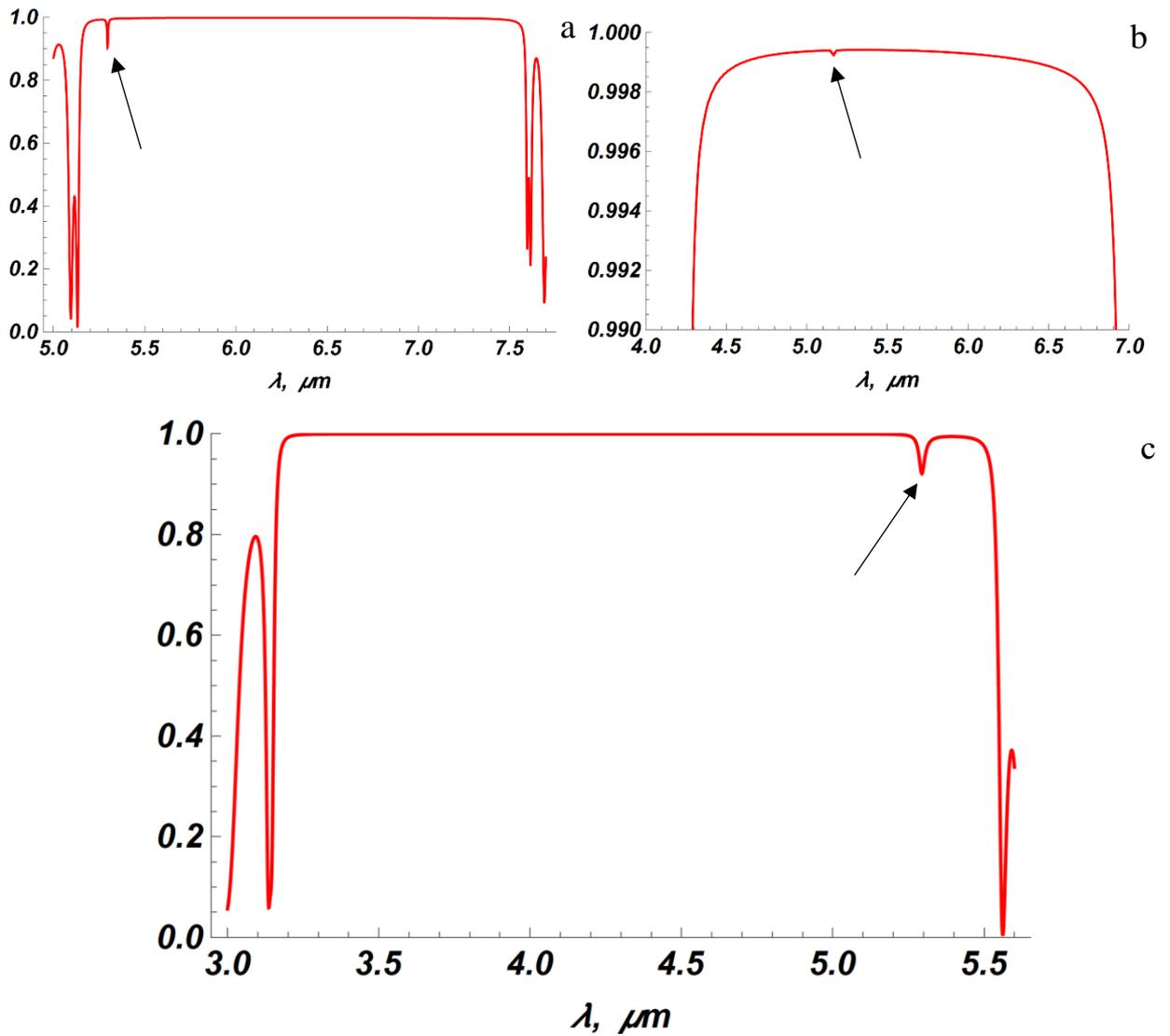

Figure 10. Reflectance spectra with defect layer with the following parameters:a) $h_d = 0.28$ μm, $\delta_i = 0$, $h_1 = 1.26$ μm, $h_2 = 0.48$ μm; b) $h_d = 0.90$ μm, $\delta_i = 0$, $h_1 = 1.00$ μm, $h_2 = 0.20$ μm; c) $h_d = 1.80$ μm, $\delta_i = 0$, $h_1 = 0.72$ μm, $h_2 = 0.275$ μm. In all cases $\lambda_{DM} = 5.30$ μm.

Let us consider each case in detail. In Figures 10a and 10c, the DM is close to the left and right boundaries of the PBG, respectively, and has a sufficient amplitude, so these options will be considered further. In Figure 10b, the DM is in the center of the PBG, and the DM does not have enough amplitude, so this option will not be considered further. The DM amplitudes in Figures 10a and 10c are almost identical. Let us consider the selected spectra when the concentration of inclusions in the defect layer changes. Figure 11 shows the dependences of the spectra R on the volume fraction δ of SARS-CoV-2 in the range from pure water ($\delta = 0$) to pure SARS-CoV-2 ($\delta = 1$) with a step of the change in the volume fraction of 0.2.

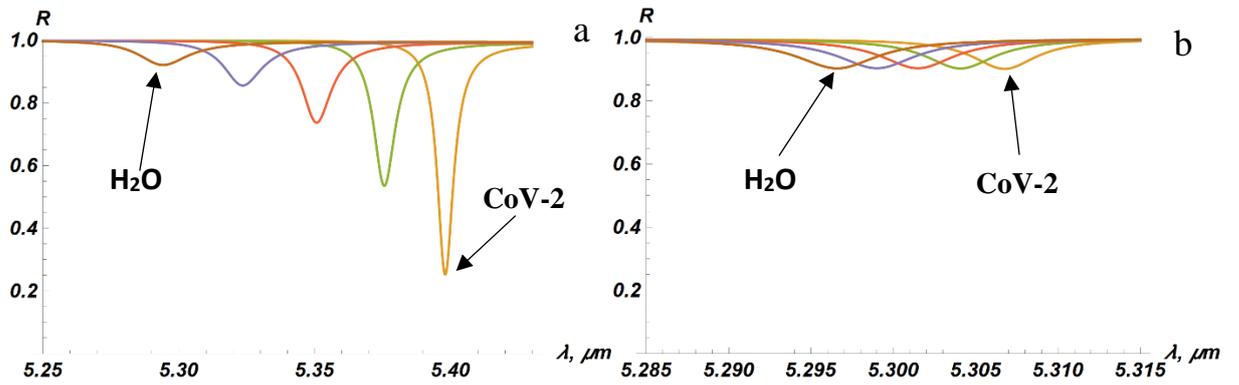

Figure 11. Spectra R with changes in concentration in increments of 0.2 at the following parameters: (a) $h_d = 0.22$ μm, $h_1 = 1.26$ μm, $h_2 = 0.48$ μm, DM is at left edge of PBG; (b) $h_d = 1.87$ μm, $h_1 = 0.72$ μm, $h_2 = 0.275$ μm, DM is at right edge of PBG.

In the spectra under consideration, as expected, there is a shift in the DM when the concentration of SARS-CoV-2 changes. In the two selected cases, the shift occurs to longer wavelengths. It is worth noting that the DM near the left edge mode of the PBG is shifted only slightly when the concentration changes, while the DM near the right edge mode of the PBG is shifted much more strongly. Another interesting fact is that the DM amplitude should increase with increasing SARS-CoV-2 concentration, as in Figure 11b, but we do not observe this effect in Figure 11a. As the pathogen concentration increases, the imaginary part of the refractive index of the defect layer decreases because the imaginary part of the SARS-CoV-2 refractive index is smaller than that of water in the selected wavelength range. On the other hand, the DM amplitude decreases as one approaches the center of the PBG. Apparently, in Figure 11a, the two described effects compensate each other, while in Figure 11b, on the contrary, the two effects are summed up, since in the two cases the direction of shift of the DM relative to the center of the PBG is different.

Let us consider the case with lower absorption; for this purpose, we will shift to the short-wave region, where the imaginary component of the refractive index of all components of our structure is close to 0. However, these spectra will be considered under the assumption that there are no SARS-CoV-2 absorption peaks in the chosen region. The reflection spectra at wavelengths $\lambda = 1$ μm and $\lambda = 2$ μm are considered. Figure 12 shows a series of R spectra with a concentration step = 0.2.

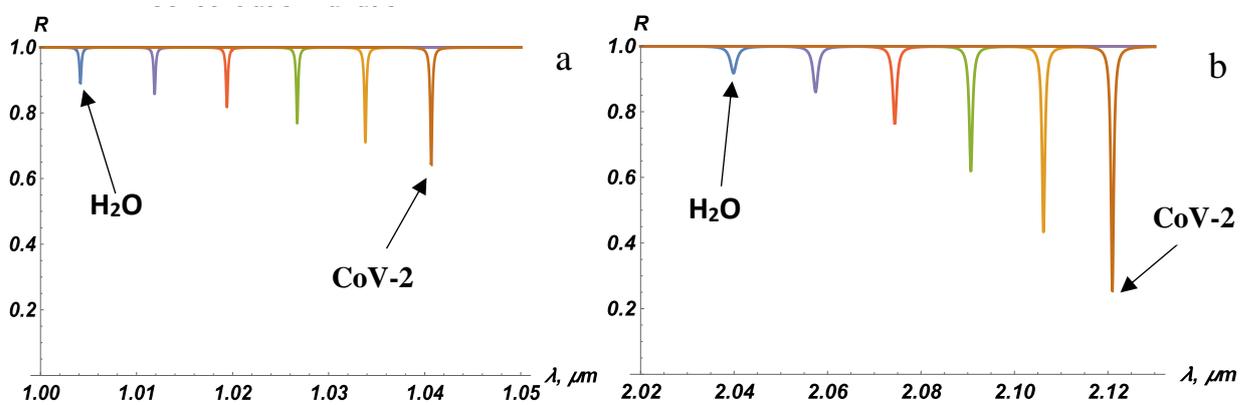

Figure 12. Spectra R with changes in concentration with the following parameters:(a) $h_d = 0.29$ μm, $h_1 = 0.14$ μm, $h_2 = 0.06$ μm; (b) $h_d = 0.630$ μm, $h_1 = 0.275$ μm, $h_2 = 0.103$ μm.

Let us now to consider the sensitivity of the DM. The characteristics of the obtained structures are shown in Figure 13 and Table 1.

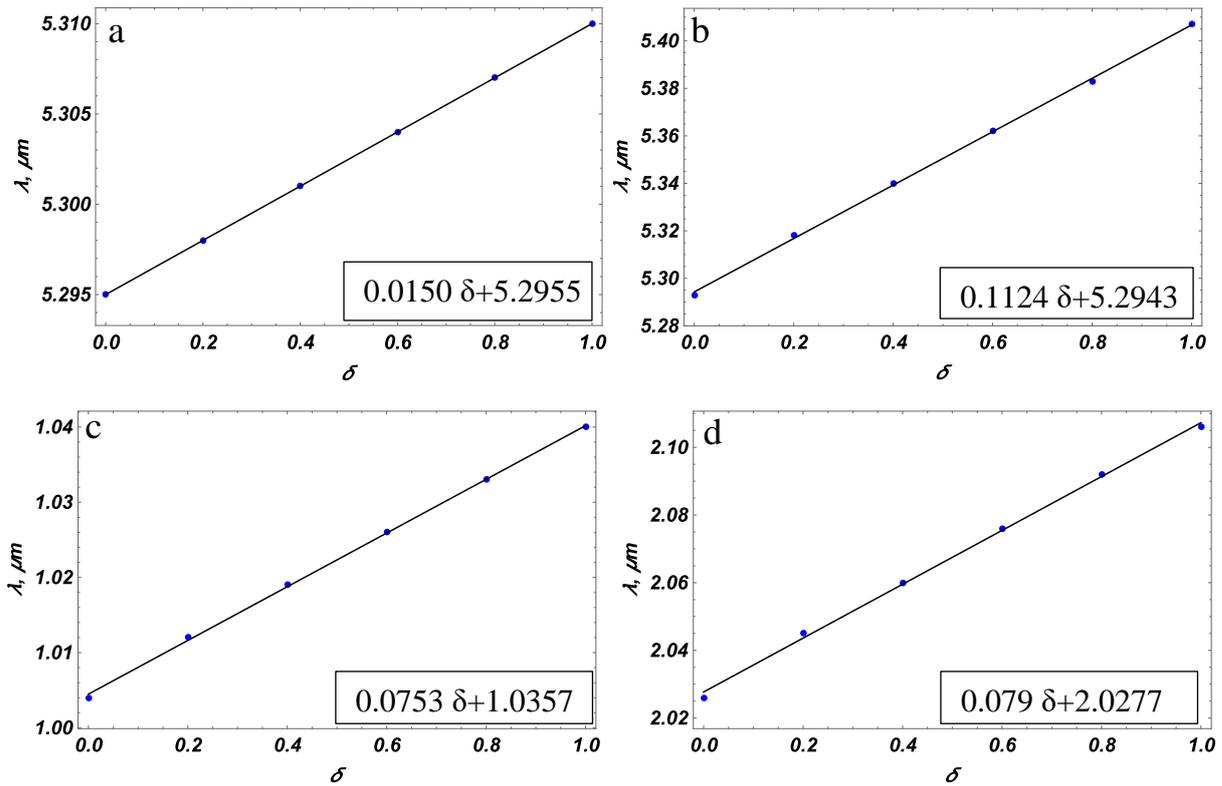

Figure 13. Dependence of the defect mode peak at different parameters: (a) $h_d = 0.22$ μm, $h_1 = 1.26$ μm, $h_2 = 0.48$ μm; (b) $h_d = 1.87$ μm, $h_1 = 0.72$ μm, $h_2 = 0.275$ μm; (c) $h_d = 0.29$ μm, $h_1 = 0.14$ μm, $h_2 = 0.06$ μm; (d) $h_d = 0.63$ μm, $h_1 = 0.275$ μm, $h_2 = 0.103$ μm.

Table 1. Comparative characteristics of sensor sensitivity at different parameters.

| № | $\lambda_{DM}$ at $\delta_i = 0$, μm, | $S1, \frac{nm}{RIU}$ | $S2, \frac{nm}{\%}$ | $\frac{S1}{\lambda_{DM}}, \frac{1}{RIU}$ | $\frac{S2}{\lambda_{DM}}, 10^{-3}, \frac{nm}{\%}$ |
|---|---|---|---|---|---|
| a | 5.295 | 102.17 | 0.15 | 0.02 | 0.03 |
| b | 5.293 | 732.43 | 1.14 | 0.14 | 0.21 |
| c | 1.004 | 272.17 | 0.36 | 0.27 | 0.35 |
| d | 2.026 | 508.20 | 0.80 | 0.25 | 0.39 |

Having analyzed the obtained data, we can come to the following conclusions:

1. The Figures 11-12 show that the DM in the short-wave region has a narrow width and high Q-factor, which allows detecting even the smallest change in the concentration of pathogens;
2. From the series of Figures 13, we can see that the dependence of the DM displacement when the concentration changes is linear at different wavelengths;
3. Comparing the results a and b, we can confirm that the sensitivity of the DM near the right PBG boundary is higher than near the left boundary;
4. Considering the sensitivity of cases b, c, d the assumption was confirmed that the sensitivity, relative to the wavelength of the DM, is higher in the short-wavelength region than in the long-wavelength region, presumably due to the fact that in the long-wavelength region we cannot get as close to the center of the PBG as in the short-wavelength region without losing the DM amplitude.

## 4. Conclusions

In conclusion, we have considered an optical biosensor capable of determining the concentration of the SARS-CoV-2 pathogen in water. Transmission and reflection spectra at different parameters of the unit cells were considered. It was found that the imaginary part of the refractive index negatively affects the amplitude of the defect mode. R and T spectra with varying thicknesses of the defect layer were considered. The spectra show that the defect modes near the long-wavelength PBG boundary have a larger amplitude in the reflection and transmission spectra.

The defect mode is stronger at the PBG boundaries, while in the center of the PBG the influence of the defect is practically not observed. A higher amplitude of the defect mode is observed in the reflection spectrum. Spectra with different concentrations of SARS-CoV-2 were analyzed and wavelength dependences of the defective mode peak on pathogen concentration were obtained. Two methods of determining the refractive index of the defective mode, the proportion method and the EMA method, were compared, no serious difference in the results obtained by the two methods was observed, in the future we plan to check other methods of calculating the refractive index depending on the concentration of the inclusions.

The optimal spectral region near the water absorption minimum, 5.3 μm, was found. A structure capable of detecting changes in pathogen concentration in this region was modeled. The considered structure has the cell parameters for the perfect PC $h_1 = 0.720$, $h_2 = 0.275$ μm and has optimal thickness of defect layer $h_d = 1.87$ μm μm (in this case there is a compromise between the CBM amplitude and the greatest distance from the edge mode). The minimum CBM amplitude for consideration was assumed to be 0.1. The sensitivity of this sensor was 1.14 nm/% and 732.43 nm/RUI. We consider this structure in the chosen range to be optimal.

According to the results of the work, it was found that to increase the sensitivity of the sensor it is necessary to reduce the imaginary component of the refractive index, so it is recommended to move to the visible region, where the imaginary component of the refractive index of water is close to zero, but for such a transition we need to measure the absorption of SARS-CoV-2 in the visible region, and select materials to create a PC with the right characteristics.

**Acknowledgments**

The work was supported by the Foundation for the Advancement of Theoretical Physics and Mathematics "BASIS" (Grant № 21-1-1-6-1).

**Conflict of interest**
No potential conflict of interest was reported by the authors.